\documentclass[12pt]{article}
\usepackage{epstopdf, graphicx}

\usepackage{times, scicite}
\usepackage{epstopdf, graphicx}

\newenvironment{sciabstract}{%
\begin{quote} \bf}
{\end{quote}}

\newcounter{lastnote}

\title{The Large, Oxygen-Rich Halos of Star-Forming Galaxies Are A Major Reservoir of Galactic Metals}

\date{}
\begin{document}
\author{
J. Tumlinson,$^{1\ast}$ 
C. Thom,$^{1}$ 
J. K. Werk,$^{2}$
J. X. Prochaska,$^{2}$
T. M. Tripp,$^{3}$\\
D. H. Weinberg,$^{4}$
M. S. Peeples,$^{5}$
J. M. O'Meara,$^{6}$
B. D. Oppenheimer,$^{7}$\\
J. D. Meiring,$^{3}$
N. S. Katz,$^{3}$
R. Dav\'{e},$^{8}$
A. B. Ford,$^{8}$
K. R. Sembach$^{1}$\\
\\
\normalsize{$^{1}$Space Telescope Science Institute, Baltimore, MD}\\
\normalsize{$^{2}$University of California Observatories-Lick Observatory, Santa Cruz, CA}\\
\normalsize{$^{3}$Department of Astronomy, University of Massachusetts, Amherst, MA}\\
\normalsize{$^{4}$Department of Astronomy, The Ohio State University, Columbus, OH}\\
\normalsize{$^{5}$Department of Physics and Astronomy, University of California, Los Angeles, CA}\\
\normalsize{$^{6}$Department of Chemistry and Physics, Saint Michael's College, Colchester, VT}\\
\normalsize{$^{7}$Leiden Observatory, Leiden University, the Netherlands}\\
\normalsize{$^{8}$Steward Observatory, University of Arizona, Tucson, AZ}\\
\\
\normalsize{$^\ast$To whom correspondence should be addressed; E-mail: tumlinson@stsci.edu.}
}

\maketitle
\begin{sciabstract}
The circumgalactic medium (CGM) is fed by galaxy outflows and accretion of intergalactic gas, but its mass, heavy element enrichment, and relation to galaxy properties are poorly constrained by observations. In a survey of the outskirts of 42 galaxies with the Cosmic Origins Spectrograph onboard the Hubble Space Telescope, we detected ubiquitous, large (150 kiloparsec) halos of ionized oxygen surrounding star-forming galaxies, but we find much less ionized oxygen around galaxies with little or no star formation. This ionized CGM contains a substantial mass of heavy elements and gas, perhaps far exceeding the reservoirs of gas in the galaxies themselves. It is a basic component of nearly all star-forming galaxies that is removed or transformed during the quenching of star formation and the transition to passive evolution.
\end{sciabstract}

Galaxies grow by accreting gas from the intergalactic medium (IGM) and converting it to stars. Stellar winds and explosions release gas enriched with heavy elements (or metals, {\it 1}), some of which is ejected in galactic-scale outflows \cite{Veilleux:05:769}. The circumgalactic medium (CGM), loosely defined as gas surrounding galaxies within their own halos of dark matter (out to 100-300 kiloparsec), lies at the nexus of accretion and outflows, but the structure of the CGM and its relation to galaxy properties are still uncertain. Galactic outflows are observed at both low ({\it 2-4}) and high ({\it 5-7}) redshift, but it is unclear how far they propagate, what level of heavy-element enrichment they possess, and whether the gas escapes the halo or eventually returns to fuel later star formation. Models of galaxy evolution require significant outflows to explain  observed galaxy masses and chemical abundances and to account for metals observed in the more diffuse IGM \cite{Springel:03:312, Oppenheimer:06:1265}. The CGM may also reflect the theoretically-predicted transition from filamentary streams of cold gas that feed low mass galaxies to hot, quasi-static envelopes that surround high mass galaxies \cite{Keres:05:2a, Dekel:06:2}.  Both outflow and accretion through the CGM may be intimately connected to the observed dichotomy between blue, star-forming, disk-dominated galaxies and red, passively evolving, elliptical galaxies with little or no star formation \cite{Kauffmann:03:33}. However, the low density of the CGM makes it extremely difficult to probe directly, thus models of its structure and influences are typically constrained indirectly by its effects on the visible portions of galaxies, not usually by observations of the gas itself.

We have undertaken a large program with the new Cosmic Origins Spectrograph (COS) aboard the Hubble Space Telescope to directly map the CGM using the technique of absorption-line spectroscopy, in which a diffuse gas is detected by its absorption of light from a background source. Our background sources are UV-bright quasi-stellar objects (QSOs), which are the luminous active nuclei of galaxies lying far behind the galaxies of interest. We focus on the ultraviolet 1032,1038 \AA\ doublet of O~VI (O$^{+5}$), the most accessible tracer of hot and/or highly ionized gas at redshift $z < 0.5$. O VI has been used to trace missing baryons in the IGM ({\it 14-17}), the association of metals with galaxies ({\it 18-20}), and coronal gas in the Milky Way halo \cite{Sembach:03:165}. 

The high sensitivity of COS enables a QSO absorption-line survey of halos around galaxies with a predetermined set of properties. We have selected 42 sample galaxies (Tables S1 and S2) that span redshifts $z_{\rm gal} = 0.10 - 0.36$, stellar masses $\log (M_*/M_{\odot}) = 9.5 - 11.5$. The QSO sightlines probe projected radial distances (i.e. impact parameters) to the galaxies of $R = 14 - 155$ kpc. We used the COS data to measure the O~VI column densities ($N_{\rm OVI}$ in cm$^{-2}$), line profiles, and velocities with respect to the target galaxies (Fig. 1, 21). We measured precise redshifts, star formation rates (SFR in $M_{\odot}$ yr$^{-1}$), and metallicities for our sample galaxies using low-resolution spectroscopy from the Keck Observatory Low-Resolution Imaging Spectrograph (LRIS) and the Las Campanas Observatory Magellan Echellette (MagE) spectrograph ({\it 22}, SOM text S2).

Our systematic sampling of galaxy properties allows us to investigate the connection between galaxies themselves and the CGM. The O~VI detections extend to $R = 150$ kpc away from the targeted galaxies, but the whole sample shows no obvious trend with radius $R$ (Fig. 2). The strong clustering of detections within $\pm$ 200 km s$^{-1}$ of the galaxy systemic velocities indicates a close physical and/or gravitational association. 

Circumgalactic medium gas as traced by O VI reflects the underlying bimodality of the general galaxy population \cite{Kauffmann:03:33, Schiminovich:07:315}. We find a correlation of $N_{\rm OVI}$ with specific star formation rate sSFR ($\equiv$ SFR/$M_*$; Fig 3). For the 30 galaxies with sSFR $\geq 10^{-11}$ yr$^{-1}$, there are 27 detections with a typical column density $\log N_{\rm OVI} = 14.5$ \cite{OVI_igm_note} and a high covering fraction $f_{\rm hit} \approx 0.8 - 1$ maintained all the way to $R = 150$ kpc (Fig. 2). For the 12 galaxies in the passive subsample (sSFR $\leq 10^{-11}$ yr$^{-1}$) there are only 4 detections with lower typical $N_{\rm OVI}$ than the star-forming subsample \cite{OVI_ell_note}. Accounting for the upper limits in $N_{\rm OVI}$ and sSFR, we can reject the null hypothesis that there is no correlation between $N_{\rm OVI}$ and sSFR at $>99.9$\% confidence for the whole sample and $>98$\% for each of the 50 kpc annuli shown in Figure 1 (SOM text S5). This effect remains even when we control for stellar mass: a Kolmogorov-Smirnov test over $\log M_* > 10.5$, where the star-forming and passive subsamples overlap, rejects at $> 99$\% confidence the null hypothesis that they draw from the same parent distribution of $N_{\rm OVI}$ (Fig. S2). We therefore conclude that the basic dichotomy between star-forming (``blue-cloud'') and passive (``red-sequence'') galaxies is strongly reflected in their gaseous halos, and that the CGM out to at least 150 kpc either directly influences or is directly affected by star formation.

O~VI is a fragile ionization state, which never exceeds a fraction $f_{\rm OVI} = 0.2$ of the total oxygen for the physical conditions of halo gas and is frequently much less abundant (Fig.~4).  Our observations imply a typical CGM oxygen mass for star-forming galaxies of 
\begin{equation}
M_{\rm O}  = 5\pi R^2 \langle N_{\rm OVI} \rangle m_{\rm O}  f_{\rm hit}
             \left(  {0.2 \over f_{\rm OVI} }  \right) = 
             1.2 \times 10^7  \left({0.2 \over f_{\rm OVI}}\right) M_\odot ~,
  \label{masseq}
\end{equation}
where we have taken a typical $\langle N_{\rm OVI} \rangle = 10^{14.5}\,{\rm cm}^{-2}$ and $R=150\,$kpc, and the hit rate correction $f_{\rm hit}$ computed separately in three 50 kpc annuli (Figs. 1 and 2). This mass of oxygen is strictly a lower limit because we have scaled to the maximum $f_{\rm OVI} = 0.2$ (Fig. 4). The corresponding total mass of circumgalactic gas is
\begin{equation}
M_{\rm gas} = 177\left({Z_\odot\over Z}\right)M_{\rm O}
            = 2\times 10^{9} \left({Z_\odot \over Z}\right)
	      \left({0.2 \over f_{\rm OVI}}\right) M_\odot ~,
  \label{totaleq}
\end{equation}
where $Z$ is the gas metallicity, and the solar oxygen abundance is $n_{\rm O}/n_{\rm H}=5\times 10^{-4}$ \cite{Asplund:09:481}. 

Even for the most conservative ionization correction ($f_{\rm OVI} = 0.2$), the OVI-traced CGM contains a mass of metals and gas that is substantial by comparison with other reservoirs of interstellar and circumgalactic gas. If our sample galaxies lie on the mean trend of gas fraction for low-$z$ galaxies \cite{Peeples:2010:3743}, they have interstellar medium (ISM) gas masses of $M_{\rm ISM} =(5-10) \times 10^9 M_{\odot}$ and contain $M^{\rm O}_{\rm ISM} = (2-10) \times 10^7 M_{\odot}$ of oxygen, taking into account the observed correlation between galaxy stellar mass and ISM metallicity (Fig. 4, SOM text S5). The minimum CGM oxygen mass is thus 10-70\% of the ISM oxygen (Figs. 4 and S4). The covering fractions and column densities we find for star-forming galaxies are insensitive to $M_*$, while the ISM metal masses decline steeply with $M_*$ according to the mass-metallicity relation. Thus the ratio of CGM metals to ISM metals appears to increase for lower mass galaxies (assuming constant $f_{\rm OVI}$), perhaps indicating that metals more easily escape from their shallower gravitational potentials. The implied total mass of circumgalactic gas $M_{\rm gas}$ is more uncertain because it can strictly take on any metallicity; for a fiducial solar metallicity, Equation~(\ref{totaleq}) implies a total CGM mass comparable to $M_{\rm ISM}$, and several times the total mass inferred for Milky Way ``high velocity clouds'' (HVCs) \cite{Putman:06:1164,Lehner:11:sub} or for low-ionization (Mg II) gas filling halos to $R = 100$ kpc \cite{Chen:10:1521}. 

For the densities typically expected at radii $R \sim 100\,$kpc,  $f_{\rm OVI}$ exceeds 0.1 only over a narrow temperature range $10^{5.4-5.6}\,$K, and it only exceeds 0.02 over $10^{5.2-5.7}\,$K (Fig.~4). Either a large fraction of CGM gas lies in this finely tuned temperature range --- a condition that is difficult to maintain because gas cooling rates peak at $T\approx 10^{5.5}\,$K --- or the CGM oxygen and gas masses are much larger than the minimum values we have quoted above.  Lower density, photoionized gas can achieve high $f_{\rm OVI} \sim 0.1$ over a wider temperature range, but at these low densities it is hard to produce a $10^{14.5}\,{\rm cm}^{-2}$ column density within the confines of a galactic halo, especially if the metallicity is low (Fig. S5). Thus $f_{\rm OVI} = 0.02$ and $Z = 0.1 Z_{\odot}$ are plausible conditions for the O VI-traced gas, but it is unlikely that both conditions hold simultaneously. However, if either one holds the CGM detected here could represent an important contribution to the cosmic budgets of metals and baryons. In either case $M_{\rm gas}$ is comparable to the total $\sim 3\times 10^{10} M_\odot$ inside $R = 300$ kpc inferred from H~I measurements at low redshift \cite{Prochaska:11:1891}, and the $\sim 4\times 10^{10} M_\odot$ inferred for the CGM surrounding rapidly star-forming galaxies at $z\approx 2-3$ \cite{Steidel:10:289}. By generalizing our typical $M_{\rm O}$ to all star-forming galaxies with $M_* > 10^{9.5} M_\odot$, we estimate that the halos of such galaxies contain $15\% \times (0.02/f_{\rm OVI})$ of the oxygen in the universe and $2\% \times (0.02 / f_{\rm OVI})\times (Z_\odot/Z)$ of the baryons in the universe. 

The metals detected out to $R \sim 150$ kpc must have been produced in galaxies, after which they were likely transported into the CGM in some form of outflow. However, these outflows need not be active at the time of observation; indeed the large masses imply long timescales. Because 1 $M_\odot$ of star formation eventually returns 0.014 $M_\odot$ of oxygen to the ISM \cite{Thomas:98:119}, at least $8.6 \times 10^8 M_\odot$ of star formation is required to yield the detected oxygen mass. This is equivalent to $\sim 3\times 10^8$ yr of star formation at the median SFR$\,= 3 \,M_\odot\,$yr$^{-1}$ of our star-forming sample, in the unlikely event that all oxygen produced is expelled to the CGM, and longer in inverse proportion to the fraction of metals retained in the ISM. Thus the detected oxygen could be the cumulative effect of steady enrichment over the preceding several Gyr, the product of sporadic flows driven by rapid starbursts and an active nucleus \cite{Tripp:PG1206}, or the fossil remains of outflows from as early as $z \sim 1.5-3$ \cite{Weiner:09:187, Steidel:10:289}. While the exact origin of the mass-metallicity relation of galaxies is not yet known, models that explain it in terms of preferential loss of metals imply that a substantial fraction of the metals produced by star formation must be ejected from the galaxy rather than retained in the ISM \cite{Peeples:2010:3743}. The CGM detected here could be a major reservoir of this ejected material, with important consequences for models of galactic chemical evolution. 

The O~VI we observe arises in bulk flows of gas over $100 - 400$ km s$^{-1}$, but the relative velocities are usually below halo escape speeds (Fig. 2), even when we take projection effects into account (Fig. S1). Thus much of the material driven into the halo by star formation could eventually be reacquired by the galaxy in ``recycled winds,'' which may be an important source of fuel for ongoing star formation \cite{Oppenheimer:10:2325}.  It is unlikely that the detected gas is predominantly fresh material accreting from the IGM because models of ``cold mode'' accretion predict very low metallicity and low covering fractions $f_{\rm hit} \sim 10-20\%$ \cite{Stewart:11:1, Fumagalli:11:2130}, and ``hot mode'' accretion typically involves gas at temperatures $T > 10^6\,$K with undetectably low $f_{\rm OVI}$. 

The passive galaxies in our sample once formed stars, thus it follows that they would once have possessed halos of ionized, metal-enriched gas visible in O~VI. The relative paucity of O~VI around these galaxies implies that this material was transformed by processes that plausibly accompany the quenching of star formation \cite{Gabor:10:749}, such as tidal stripping in group environments, re-accretion onto the galaxy in ionized form, or heating or cooling to a temperature at which O~VI is too rare to detect. Our findings present a quantitative challenge for theoretical models of galaxy growth and feedback, which must explain both the ubiquitous presence of massive, metal-enriched ionized halos around star-forming galaxies and the fate of these metals after star formation ends.

\bibliographystyle{Science}

{\bf Acknowledgements:} We thank the anonymous reviewers for constructive comments. Based on observations made for program GO11598 with the NASA/ESA {\it Hubble Space Telescope}, obtained at the Space Telescope Science Institute, operated by AURA under NASA contract NAS 5-26555, and at the W.M. Keck Observatory, operated as a scientific partnership of the California Institute of Technology, the University of California and NASA. The Observatory was made possible by the generous financial support of the W.M. Keck Foundation. The {\it Hubble} data is available from the MAST archive at http://archive.stsci.edu. M. S. P. acknowledges support from the Southern California Center for Galaxy Evolution, a multi-campus research program funded by the UC Office of Research. \\

\noindent {\bf Supporting Online Material:} \\
SOM Text\\
Figs. S1 to S5 \\
Tables S1 and S2 \\
References (40-62)\\

\clearpage
\begin{figure*}[!ht]
\begin{center}
\includegraphics[width=6in]{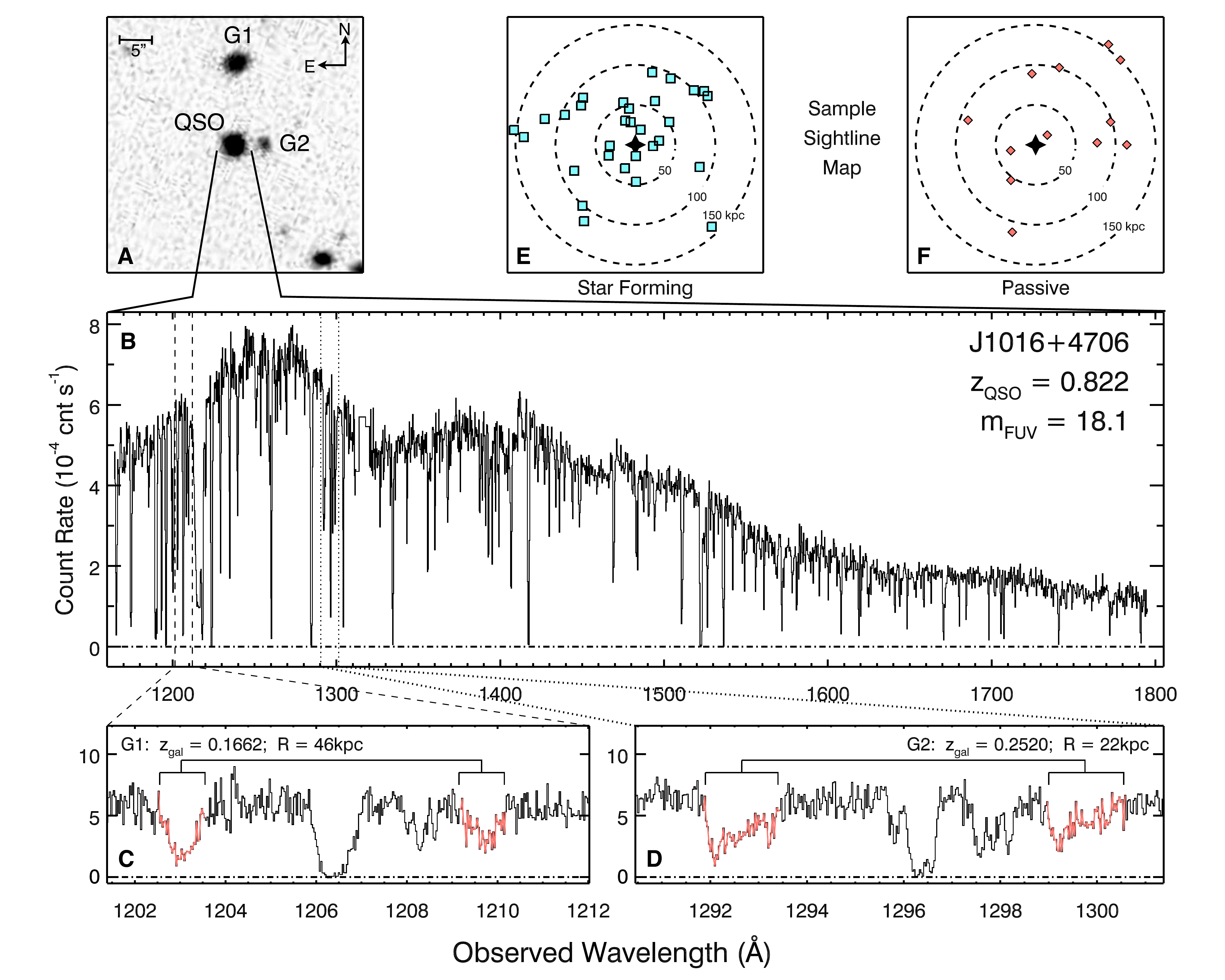}
\end{center}
\label{datafig}
\end{figure*}
{\bf Figure 1:} An illustration of our sampling technique and data. ({\bf A}) An SDSS composite image of the field around the QSO J1016+4706 with two targeted galaxies, labeled G1 and G2, which are both in the star-forming subsample. ({\bf B}) The complete COS count-rate spectrum (counts s$^{-1}$) versus observed wavelength. The panels below concentrate on the redshifted O~VI 1032,1038 \AA\ doublet for galaxies G1 ({\bf C}) and G2 ({\bf D}).  The upper right panels illustrate the full sample by showing the locations of all sightlines in position angle and impact parameter $R$ with respect to the targeted galaxies, for the star-forming ({\bf E}) and passively evolving ({\bf F}) subsamples. The circles mark $R = $ 50, 100, and 150 kpc.

\clearpage
\begin{figure}[!ht]
\begin{center}
\includegraphics[width=3.7in]{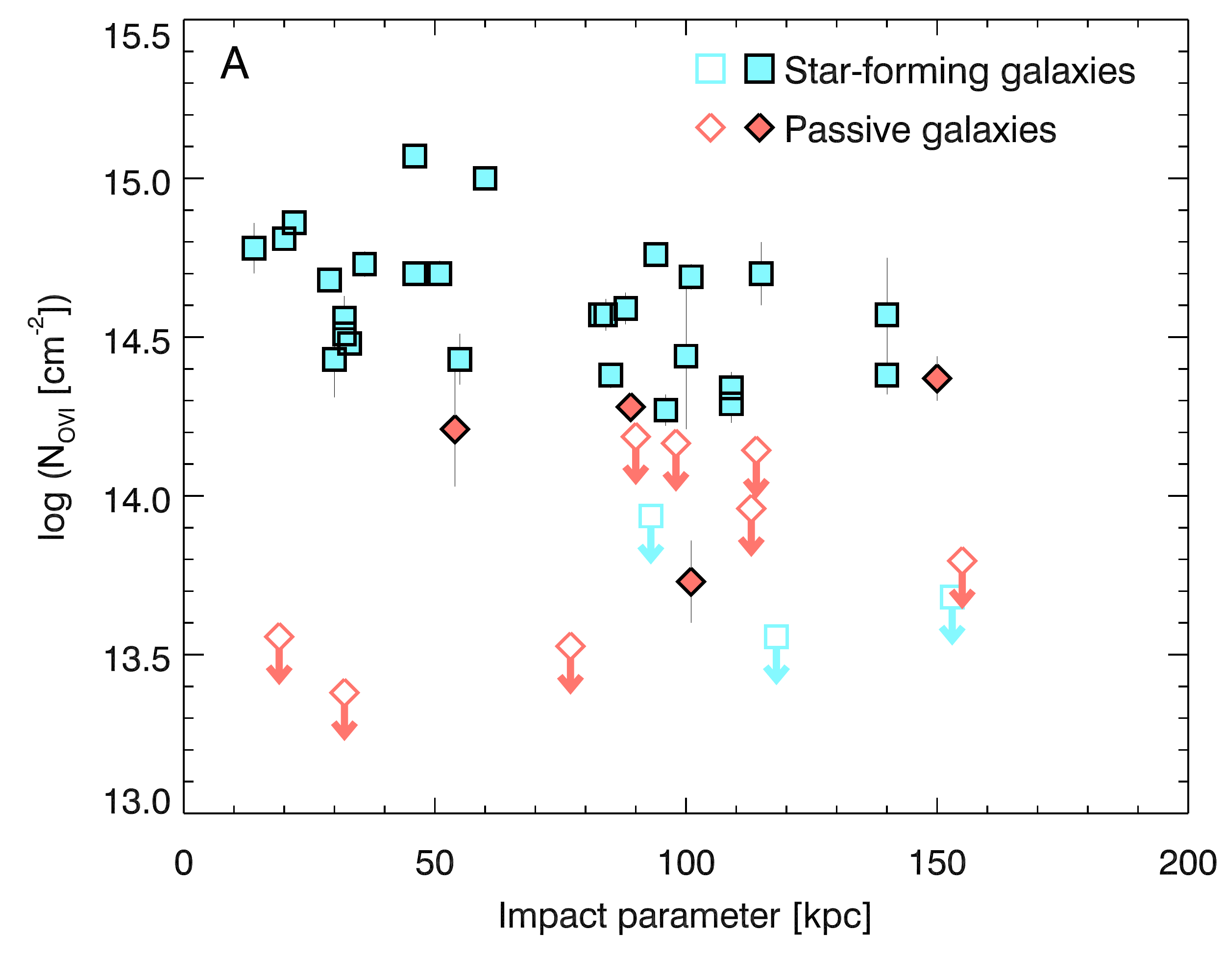} 
\includegraphics[width=3.7in]{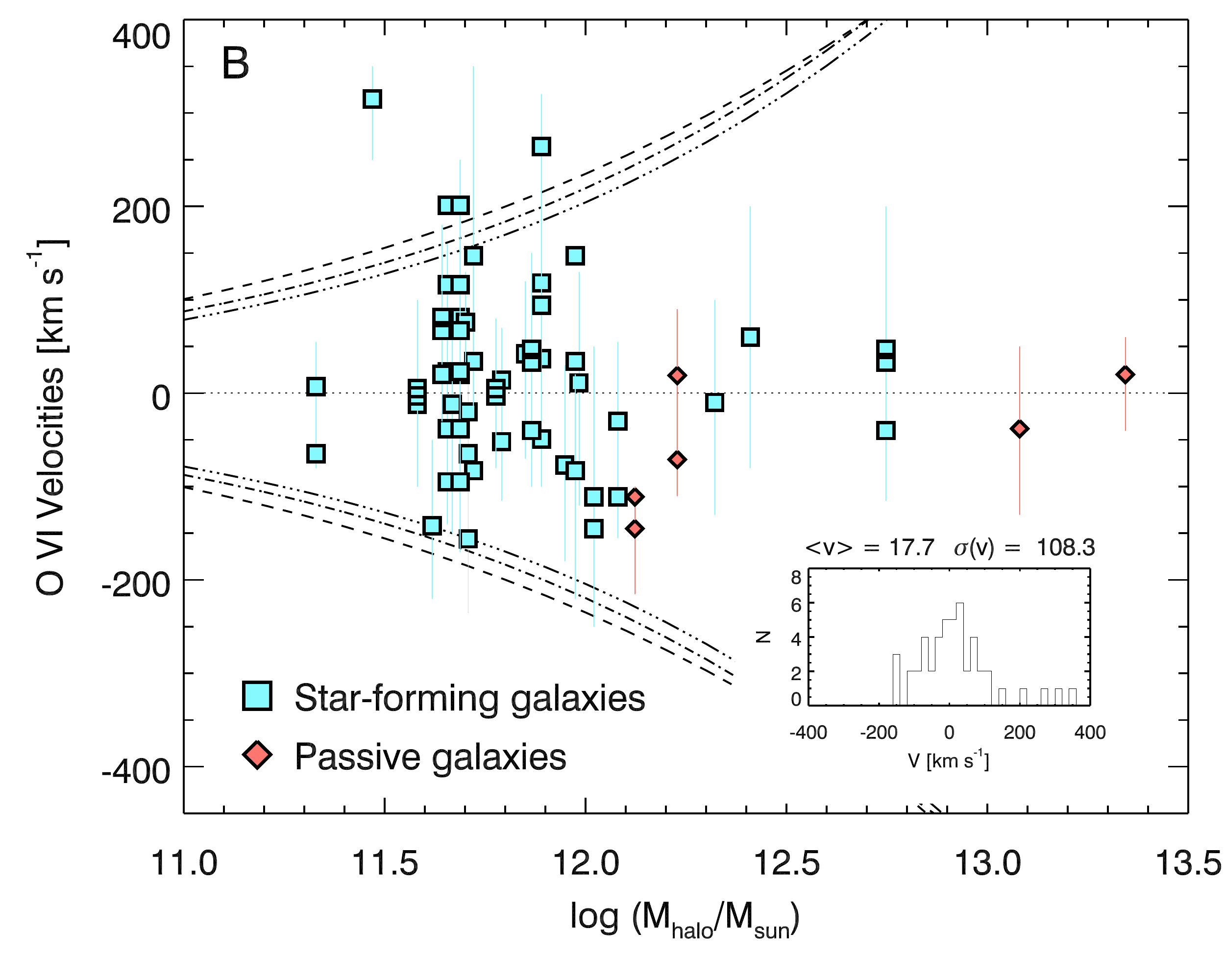} 
\end{center}
\label{fig_colden_impactpar}
\end{figure}
{\bf Figure 2:} O~VI association with galaxies. ({\bf A}) O~VI column density, $N_{\rm OVI}$ in cm$^{-2}$, vs. $R$ in kiloparsec for the star-forming (blue) and passive (red) subsamples. Filled and open symbols mark O~VI detections and $3\sigma$ upper limits, respectively. The detections in the star-forming galaxies maintain $\log N_{\rm OVI} \sim 14.5$ to $R \sim 150$ kpc, the outer limit of our survey. ({\bf B}) Component centroid velocities with respect to galaxy systemic redshift for O~VI detections, versus inferred dark-matter halo mass. The range bars mark the full range of O~VI absorption for each system. The inset shows a histogram of the component velocities. The dashed lines mark the mass-dependent escape velocity at $R = 50$, $100$, and $150$ kpc from outside to inside.

\clearpage
\begin{figure*}[!ht]
\begin{center}
\includegraphics[width=3.8in]{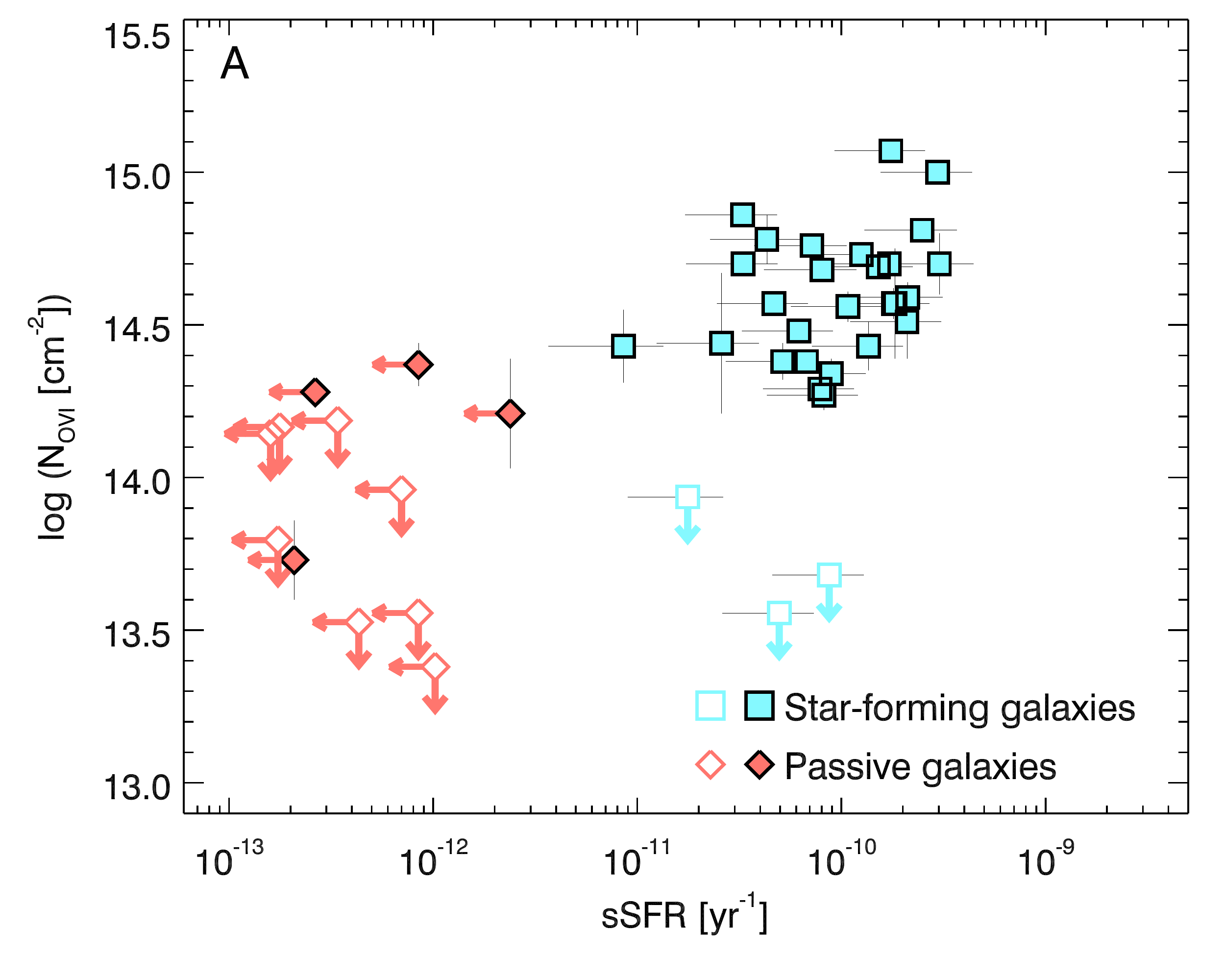} 
\includegraphics[width=3.8in]{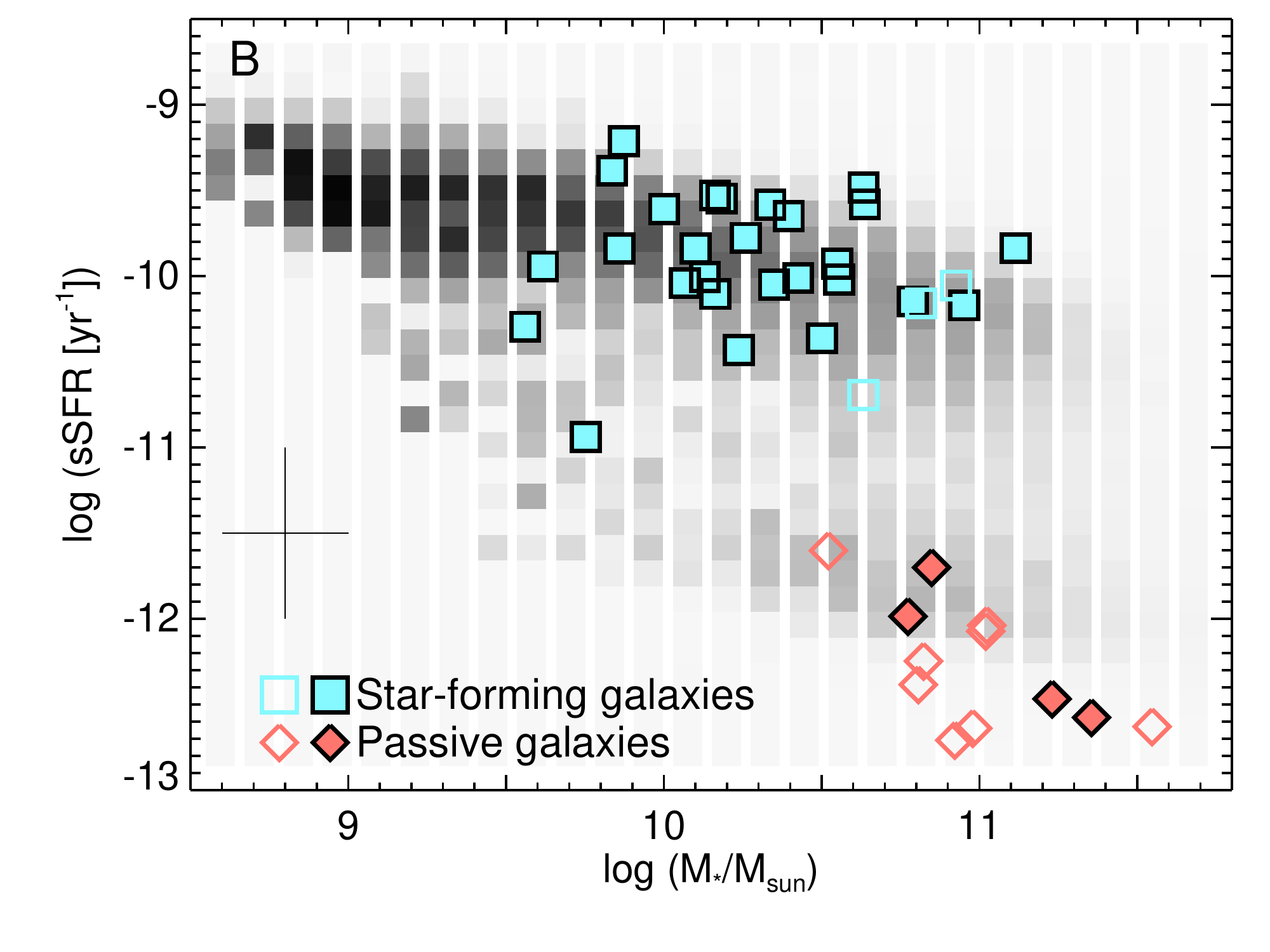}
\end{center}
\label{fig_colden_ssfr_rcuts}
\end{figure*}
{\bf Figure 3:} O~VI correlation with galaxy properties. ({\bf A}) O~VI column density versus sSFR ($\equiv M_*/SFR$). Star-forming galaxies are divided from passively evolving galaxies by sSFR $\approx 10^{-11}$ yr$^{-1}$, our detection limit is sSFR $\sim 5\times 10^{-12}$ yr$^{-1}$). ({\bf B}) The galaxy color-magnitude diagram (sSFR versus $M_*$) for SDSS+GALEX galaxies from \cite{Schiminovich:07:315}.  

\clearpage
\begin{figure*}
\begin{center}
\includegraphics[width=6.2in]{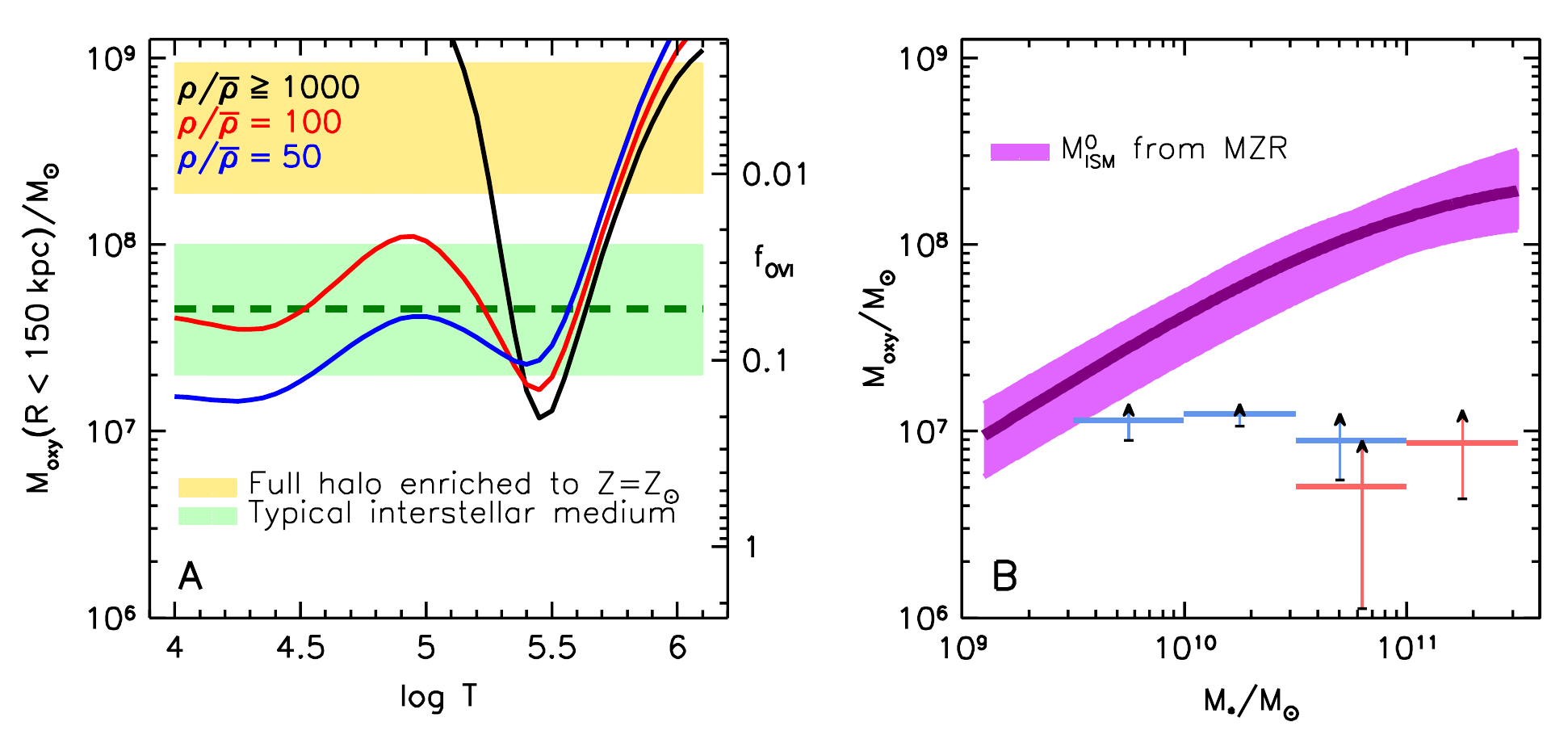}
\end{center}
\label{ovifracfig}
\end{figure*}
\vspace{-0.20in}
{\bf Figure 4:} CGM oxygen masses compared to galactic reservoirs. ({\bf A}) The curves and the axis labels at right show the fraction of gas-phase oxygen in the O~VI ionization state $f_{\rm OVI}$ as a function of temperature, for three overdensities relative to the cosmic mean, $\rho/\bar{\rho}$.  All $\rho/\bar{\rho} \geq 1000$ track the black curve on which collisional ionization dominates, while for lower values photoionization by the extragalactic background can increase $f_{\rm OVI}$ at low $T$. For gas that traces dark matter, $\rho/\bar{\rho}=1000$ is typical at $R\approx 100$ kpc, while $\rho/\bar{\rho}=50-100$ for the outskirts of the halo. The pale green band shows the expected oxygen mass of the galaxies' ISM if they lie on the standard relation between $M_{\rm ISM}$ and $M_*$ and follow the mass-metallicity relation (MZR). The green dashed line shows the oxygen mass produced by $3 \times 10^{9} M_{\odot}$ of star formation. The yellow band shows the expected oxygen mass for the extreme assumption that the typical host dark-matter halos ($2\times 10^{11} - 10^{12} M_\odot$) have the universal baryon fraction and solar metallicity. ({\bf B}) The CGM oxygen masses compared with the interstellar oxygen mass as a function of $M_*$. Points with range bars show the CGM oxygen mass $M_{\rm O}$ implied by Equation (1) for $f_{\rm OVI} = 0.2$, calculated separately for star-forming (blue) and passive (red) galaxies according to the hit rates in four bins of stellar mass. The purple curves show the calculated $M_{\rm ISM}^{\rm O}$ for typical star forming galaxies in the SDSS, accounting for the mean MZR in the central curve and its uncertainties in the shaded region. The data points increase their mass in inverse proportion to $f_{\rm OVI}$.

\clearpage

\begin{center} 
{\Large \bf Supporting Online Material} 
\vspace{0.2in} 
\\{ \Large \bf The Large, Oxygen-Rich Halos of Star-Forming Galaxies Are A Major Reservoir of Galactic Metals} \\
\vspace{0.2in} 

J. Tumlinson, C. Thom, J. K. Werk, J. X. Prochaska, T. M. Tripp, D. H. Weinberg, M. S. Peeples, J. M. O'Meara, B. D. Oppenheimer, J. D. Meiring, N. S. Katz, R. Dav\'{e}, A. B. Ford, \& K. R. Sembach
\end{center}

\noindent{\Large \bf S1 \,\, Survey Design and Data Collection} \\

In recent years, the joint advances of the SDSS DR5 QSO catalog \cite{Schneider:07:102}, the GALEX all-sky UV survey \cite{gr4ref}, and the installation of the Cosmic Origins Spectrograph \cite{Osterman:11:157} on-board {\it HST} have increased the numbers of UV-bright QSOs that can be observed with HST by an order of magnitude or more. We can now select a sample of QSOs whose sightlines pierce the halos of foreground galaxies with a predetermined range of properties,  thus establishing a uniform sampling of galaxy stellar mass ($M_{*}$), star-formation rate (SFR), and impact parameter ($R$).

We began our sample selection by choosing QSOs at $z < 1$ that were bright enough to obtain COS spectra with S/N $\sim  8-10$ in five HST orbits or less. We require $z_{\rm gal} > 0.11$ to place the O VI doublet within the $\lambda > 1140 $ \AA\ wavelength range of COS. From the list of approximately 1000 such candidates that met all these criteria, we constructed an HST program of 39 QSOs and 134 orbits, with the galaxies sampling the range $\log M_*/M_{\odot} = 10 - 11$ (with some scatter out of this original range caused by errors in photometric redshifts) and $R = 0 - 150$ kpc as uniformly as possible. 

We obtained COS FUV spectroscopy of this sample of QSOs between March and October 2010. The spectra were obtained at a resolving power of $\Delta \lambda / \lambda \sim$ 18,000$-$20,000 over $1140 - 1750$ \AA\ using the G130M and G160M gratings of COS. After standard processing by the CALCOS pipeline (v2.12), the individual exposures were co-added in counts space to provide a final, reduced 1D spectrum trace (using a method detailed by \cite{Meiring:11:35, Tumlinson:11:111}). 

Out of the 39 program targets, six COS spectra either did not provide coverage of the O VI doublet with data of acceptable quality, or they require special processing that is not yet implemented within CALCOS. We fully analyzed a total of 33 program datasets for this study. Our final sample consisted of 42 galaxies in the foreground of these QSOs: 30 star-forming and 12 passively evolving galaxies. Eleven of the galaxies were not part of the original selection, although they meet the original criteria. These ``bonus'' galaxies were confirmed  at different redshifts along the line of sight by optical spectroscopy, and we have checked that  these galaxies do not bias the overall sample in any of the galaxy properties. A full presentation of the galaxy survey technique, the optical spectroscopy, and the associated errors is presented by ({\it 22}). We exclude from the present sample those galaxies that did not provide good coverage of O VI in the COS data; the 42 galaxies in the present sample are listed in Tables S1 and S2 for star-forming and passive galaxies, respectively. We label our galaxies with two numbers: the position angle from the QSO, N through E, and the angular separation in arcseconds. The QSO and this galaxy designation can be used to cross-reference to the tables of the larger database in ({\it 22}). 

To measure the precise redshifts and star formation rates of the galaxies, we obtained spectra of the galaxies with the Low-Resolution Imaging Spectrometer (LRIS) on the Keck telescope, and the Magellan Echellette (MagE) spectrograph on the Magellan Clay Telescope. With LRIS, we employed a 1$''$ slit and the D560 dichroic, using the 600/4000 l/mm grism (blue side) and 600/7500 l/mm grating (red side). This setup resulted in spectral coverage between 3000 and 5500 \AA\ (blue side), and 5600 to 8200 \AA\ (red side). On the blue side, binning the data 2 $\times$ 2 resulted in a dispersion of 1.2 \AA\ per pixel and a FWHM resolution of $\sim$280 km/s. On the red side, the data were binned 1 $\times$ 2, resulting in a dispersion of 2.3 \AA\ per pixel and a FWHM resolution of $\sim$200 km/s. Exposure times were typically 800s in the blue and 2 $\times$ 360 s in the red, which resulted in signal-to-noise ratios of at least 3 per pixel for strong nebular emission lines in the galaxy spectra. Data reduction and calibration were carried out using the LowRedux IDL software package \cite{LowRedux}, which includes flat fielding to correct for pixel-to-pixel response variations and larger scale illumination variations, wavelength calibration, sky subtraction, and an initial flux calibration using the spectrophotometric standard star G191B2B.  

\clearpage

{\small 
\begin{center} 
Table S1: O VI Sample Galaxies (Star-forming) 
 \begin{tabular*}{0.60\textwidth}{@{\extracolsep{\fill}} ccc}
  \hline
QSO & Galaxy & $N_{\rm O VI}$ cm$^{-2}$ \\
  \hline
J0042-1037  &  358\_9  &  $14.78 \pm  0.08$ \\ 
J0401-0540  &  67\_24  &  $14.57 \pm  0.03$ \\ 
J0820+2334  &  260\_17  &  $14.43 \pm  0.12$ \\ 
J0910+1014  &  34\_46  &  $14.70 \pm  0.10$ \\ 
J0914+2823  &  41\_27  &  $14.69 \pm  0.04$ \\ 
J0925+4004  &  193\_25  &  $<13.94$ \\ 
J0943+0531  &  106\_34  &  $<13.56$ \\ 
J0943+0531  &  227\_19  &  $14.44 \pm  0.23$ \\ 
J1009+0713  &  170\_9  &  $15.07 \pm  0.02$ \\ 
J1009+0713  &  204\_17  &  $15.00 \pm  0.03$ \\ 
J1016+4706  &  274\_6  &  $14.86 \pm  0.02$ \\ 
J1016+4706  &  359\_16  &  $14.70 \pm  0.03$ \\ 
J1112+3539  &  236\_14  &  $14.70 \pm  0.04$ \\ 
J1133+0327  &  164\_21  &  $14.43 \pm  0.08$ \\ 
J1233+4758  &  94\_38  &  $14.38 \pm  0.06$ \\ 
J1233-0031  &  168\_7  &  $14.68 \pm  0.03$ \\ 
J1241+5721  &  199\_6  &  $14.81 \pm  0.02$ \\ 
J1241+5721  &  208\_27  &  $14.76 \pm  0.02$ \\ 
J1245+3356  &  236\_36  &  $14.34 \pm  0.05$ \\ 
J1330+2813  &  289\_28  &  $14.38 \pm  0.04$ \\ 
J1342-0053  &  157\_10  &  $14.48 \pm  0.03$ \\ 
J1419+4207  &  132\_30  &  $14.59 \pm  0.05$ \\ 
J1435+3604  &  68\_12  &  $14.73 \pm  0.04$ \\ 
J1435+3604  &  126\_21  &  $14.57 \pm  0.05$ \\ 
J1437+5045  &  24\_13  &  $14.51 \pm  0.12$ \\ 
J1437+5045  &  317\_38  &  $14.57 \pm  0.18$ \\ 
J1445+3428  &  232\_33  &  $14.29 \pm  0.06$ \\ 
J1550+4001  &  97\_33  &  $<13.68$ \\ 
J1555+3628  &  88\_11  &  $14.56 \pm  0.05$ \\ 
J1619+3342  &  113\_40  &  $14.27 \pm  0.05$ \\ 
  \hline
\end{tabular*}

\end{center} }

\clearpage

{\small 
\begin{center} 
Table S2: O VI Sample Galaxies (Passive) 
 \begin{tabular*}{0.60\textwidth}{@{\extracolsep{\fill}} ccc}
  \hline
QSO & Galaxy & $N_{\rm O VI}$ cm$^{-2}$ \\
  \hline
J0226+0015  &  268\_22  &  $<13.53$ \\ 
J0910+1014  &  35\_14  &  $14.21 \pm  0.18$ \\ 
J0928+6025  &  110\_35  &  $<14.19$ \\ 
J0935+0204  &  15\_28  &  $<13.96$ \\ 
J0943+0531  &  216\_61  &  $<13.80$ \\ 
J0950+4831  &  177\_27  &  $14.28 \pm  0.04$ \\ 
J1157-0022  &  230\_7  &  $<13.56$ \\ 
J1220+3853  &  225\_38  &  $14.37 \pm  0.07$ \\ 
J1342-0053  &  77\_10  &  $<13.38$ \\ 
J1550+4001  &  197\_23  &  $13.73 \pm  0.13$ \\ 
J1617+0638  &  253\_39  &  $<14.17$ \\ 
J2257+1340  &  270\_40  &  $<14.14$ \\ 
  \hline
\end{tabular*}
\end{center} }

The MagE spectrograph provides moderate resolution spectral coverage between 3200 \AA~ and 10000 \AA. These data were acquired with a 0.7$''$ slit and binned 1 $\times$ 1, giving typical dispersions of 0.3 \AA\ per pixel at 4500\AA, and 0.5 \AA\ per pixel at 7800 \AA.  The average emission-line FWHM is 55 km s$^{-1}$.  Exposure times varied between 600 and 1200 seconds, depending on the target galaxy apparent magnitude. The MagE spectra were reduced using the MASE pipeline \cite{MASE}. One-dimensional spectra are optimally extracted from the 2D reduced images. Spectrophotometric standard stars taken at a variety of airmasses were used to  flux-calibrate the data.

We make three corrections to the flux scale of our optical spectra: 1.  A correction for slit losses that brings the spectra to an absolute flux scale. 2. A correction for Galactic extinction, assuming the $E(B-V)$ value from \cite{Schlegel:98:525} and a Galactic extinction law \cite{Cardelli:89:245}. 3. A correction for interstellar reddening to all line measurements from the observed H$\gamma$ to H$\beta$  and H$\alpha$ to H$\beta$ ratios. To bring the LRISb spectra to an absolute flux scale, we convolve it with the SDSS $g$-band filter response curve and scale the flux of the blue spectrum to match the reported SDSS-DR7 petrosian magnitude.  Similarly, we match the $i$-band magnitude for our LRISr spectra. The median values of these two flux factors are 1.94 (blue) and 1.77 (red), corresponding to median slit light losses of 48\% and 43\%. These corrections, which provide accurate (at least as accurate as SDSS photometry) absolute fluxing,  account for both slit losses and red/blue side matching in a self-consistent global fit. Furthermore, when performing Balmer corrections, we use as many Balmer lines as are detected in the spectrum and find generally good agreement after we have corrected the red and blue sides for slit losses and differences in absolute flux scales on the red and blue sides. However, sometimes there are only two Balmer lines detected, and occasionally, only one. For Balmer emission-line measurements, we account for contamination from underlying stellar absorption by fitting the continuum in the trough of detectable absorption. The overall effect on the line flux of the Balmer absorption ranges from 10\% to 60\% for the H$\beta$ emission line (when absorption is apparent).  As for the LRIS data, we account for slit-losses by scaling the spectra to match SDSS $r$-band photometry (MagE contains no dichroic, so only a single band is used).
\\

\noindent{\Large \bf S2 \,\, Galaxy Properties from Spectroscopy} \\

Precise and accurate systemic redshifts were obtained for galaxies using a modified version of the SDSS IDL code ``zfind'', which works by fitting smoothed template SDSS eigenspectra to the galaxy emission-line spectra on both the red and blue sides. To account for systematic uncertainties in the absolute wavelength calibration and instrument flexure during the LRIS exposures, we adopt a conservative 30 km s$^{-1}$ error on the galaxy redshifts. 

To obtain an estimate of the stellar mass of each galaxy, we used version 4\_2 of the {\it kcorrect} IDL package \cite{Blanton:07:734}, the SDSS DR7 galactic reddening corrected, asinh {\emph{ugriz}} magnitudes, and the {\emph{zfind}} spectroscopic redshifts. The stellar masses output by the routine {\it{sdss\_kcorrect}} are computed using a Chabrier IMF with the 1994 Padova isochrones.  We raise the masses by a factor of 1.65 (0.22 dex) to compare them with quantities determined using a Salpeter IMF (i.e. SFRs, described below) and assign errors of 0.2 dex on $M_*$ propagated from the SDSS photometry and the systematic errors in the adopted mass conversion \cite{Bell:03:289, McIntosh:08:1537}. The masses we adopt here have been corrected assuming the 5-year WMAP cosmology with a Hubble parameter $h=0.72$ \cite{Dunkley:09:306}. 

We then calculate a current SFR using the Balmer emission lines H$\alpha$ and H$\beta$.  For the former, we use the calibration of \cite{Kennicutt:98:541} where SFR [M$_{\odot}$ yr$^{-1}$] $=$ 7.9 $\times$ 10$^{-42}$ L$_{H\alpha}$ [ergs s$^{-1}$].  This SFR calibration is derived from stellar population synthesis models that assume a Salpeter IMF and solar metallicity.  When H$\alpha$ is not observed in a spectrum, we use the same SFR calibration \cite{Kennicutt:98:541} divided by a factor of 2.86 for the H$\beta$ emission line. This factor of 2.86 is the intrinsic ratio of  H$\alpha$/H$\beta$ at an effective temperature of 10,000 K and electron density of 100 cm$^{-3}$ for Case B recombination \cite{Hummer:87:801}. Thus, the SFRs derived from H$\alpha$ and H$\beta$ are identical when we calculate a dust correction that gives H$\alpha$/H$\beta$ = 2.86. The SFRs used in comparisons in this paper adopt the measurement from the best available line, or both as the case may be. 

When a galaxy's spectrum contains no emission lines, we measure an upper limit to the SFR by measuring the boxcar noise at the positions of  H$\beta$ and H$\alpha$. We use 3$\sigma$ line flux limits as our SFR upper limits in these cases, approximately 1/3  of our sample.  We adopt conservative 3$\sigma$ limits to the SFRs since in these galaxies we are unable to make a correction for dust. \\

\noindent{\Large \bf S3 \,\, O VI Line Identification and Classification} \\

We begin by shifting the reduced, co-added COS spectra to the rest-frame of the target galaxy, setting $v = 0$ to its measured systemic redshift. We then generate apparent optical depth plots \cite{Savage:91:245} for the two lines of the O VI doublet to simplify their identification and confirmation, and to easily identify contaminating absorption. We then inspect the data to look for O VI absorption within $\pm 600$ km s$^{-1}$ of the systemic redshift. Most of the O VI absorbers are readily apparent and easily confirmed in both lines of the doublet, but the redshifts of some targeted systems place one line of the doublet where it cannot be detected, either off the 1140 \AA\ short end of the COS detector ($z_{\rm OVI} < 0.1$), or underneath Galactic Ly$\alpha$ ($z_{\rm OVI} \sim 0.17$) or geocoronal O I airglow emission ($z_{\rm OVI} \sim 0.25$).  

Contaminating absorption from other intervening absorption-line systems, or from the foreground ISM of the Milky Way, is always a risk. We attempt to positively identify every detected feature within the $\pm 600$ km s$^{-1}$ range at the position of O VI. Most such features not associated with the targeted galaxy are positively identified HI or metal lines from unaffiliated absorbers at other redshifts along the sightline. We generally confirm the presence of an O VI absorber by detecting the doublet in the expected 2:1 absorption ratio, by excluding alternative identifications for the line (such as another ionization species from another cataloged absorption-line system along the sightline), and by requiring that the O VI is aligned to within reasonable limits with other species in its absorption system. In the few cases where the data do not cover both lines of the doublet because of detector edges or contaminating Galactic absorption, the doublet ratio is not available but the latter two criteria must be satisfied to count as a detection. 

Most contaminating absorption that directly impacts the detected O VI is easily accounted for in profile fits as nuisance absorption because it affects one line of the doublet but not both. After all these steps to identify and/or account for contaminating absorption, we conclude that any residual contamination to the detected O VI is of negligible impact to our statistical results. We confirm non-detections or ``misses'' with at least one line of the doublet. For absorbers with both line of the doublet detected and uncontaminated by unrelated absorption, we have found that the optical depths of the two profiles agree to within errors over the  full range of velocity, which indicates that partial covering of the background source and line saturation are not important effects on our measurements.  \\

\noindent{\Large \bf S4 \,\, O VI Column Densities and Kinematics} \\

We obtain column density and kinematic estimates using both direct integration and profile-fitting techniques, which complement one another and reduce total systematic error. Each line of each O VI absorber has its total column density $N_{\rm OVI}$ and the corresponding errors obtained by direct integration \cite{Savage:91:245} over velocity limits that were visually determined for each system and applied to both lines of the doublet. For non-detections, $3\sigma$ upper limits to the line strength are obtained from estimates of the RMS noise in the continuum where the line is not detected, typically over the range $\pm 50$ km s$^{-1}$ around the corresponding Lyman series lines of the absorption system, or around the galaxy systemic redshift, in cases where no Ly$\alpha$ absorption is detected. 

Profile fitting improves on direct line integration of column densities using information about the line shapes and placement to constrain the fit. It also helps assess line saturation by taking into account the true line-spread function (LSF) of the instrument. While fitting can give good estimates of column density, intrinsic linewidth, and component structure, the derived fits can be degenerate and/or non-unique in cases with closely blended components. Nevertheless the systematic effects that complicate profile fitting are different from those that affect direct integration, so we use both techniques to assess line saturation and minimize total systematic errors. 

Our profile-fitting procedure goes as follows. We extract from each reduced COS spectrum two short slices of $\pm$600 km s$^{-1}$ around each target systemic redshift, at the position of the two  O VI lines. The line profiles are fitted using Voigt profiles with parameters $N$, $v$, and $b$ obtained by $\chi ^2$ minimization. Most well-detected O VI systems have the fit parameters tied together for both lines of the doublet; this tying is also done separately for each velocity component where there is more than one. The lines are not tied together where one is unavailable or too contaminated to lead to a good fit. Nuisance and/or contaminating lines are fitted where necessary (usually with arbitrary atomic data). The O~VI absorbers generally require $1-4$ velocity components to generate a good fit, with roughly two-thirds of the sample showing only one component at COS resolution, about one-fifth showing two components, and the remaining systems three or four components. The fitting uses the actual COS LSF \cite{Ghavamian:09:1} based on the fitted line's {\it observed} wavelength and convolves the model spectrum with this LSF before comparing the model to the data. The result of this fitting process is a set of fitted parameters ($N$, $b$, $v$) for each component in each system, and their errors derived from the parameter covariance matrix around the best-fit point. The statistical errors on $N_{\rm OVI}$ are generally $<0.1$ dex. We find that the saturation corrections implied by the difference between direct integration and the profile-fitted column densities are small, with the fitted values generally being larger by $< 0.1- 0.2$ dex (which also lends confidence to the errors estimated from covariance of the fit parameters). We adopt the fitted values in the main analysis. If the saturation correction is underestimated at COS resolution then the $N_{\rm OVI}$ generally increase and our main conclusions get stronger. 

The strong O VI we detect is in close velocity association with the galaxies. Figure~2 of the main text shows the  component velocity centroids and the full range of the O~VI absorption with respect to the galaxy systemic redshift. The errors on the component velocities are generally $\pm\, 30$ km s$^{-1}$ and arise from a combination of the $25$ km s$^{-1}$ in the galaxy systemic redshift and $10-15$ km s$^{-1}$ errors in the COS wavelength solution; multiple components in the same system would shift together owing to errors in either one of these quantities. The distribution of the detected component velocities has a mean of 18 km s$^{-1}$ and a standard deviation of 108 km s$^{-1}$, and so the strong O VI we detect is closely associated with the sample galaxies in velocity space. This is perhaps not unexpected, given that we have deliberately probed close to galaxies, but there is nothing in the close physical separation from the sightlines that requires the O VI to favor the galaxy systemic velocity, as we find, over the broader range over which we conducted our search. 

\begin{figure*}[!ht]
\begin{center} 
\includegraphics[width=4.5in]{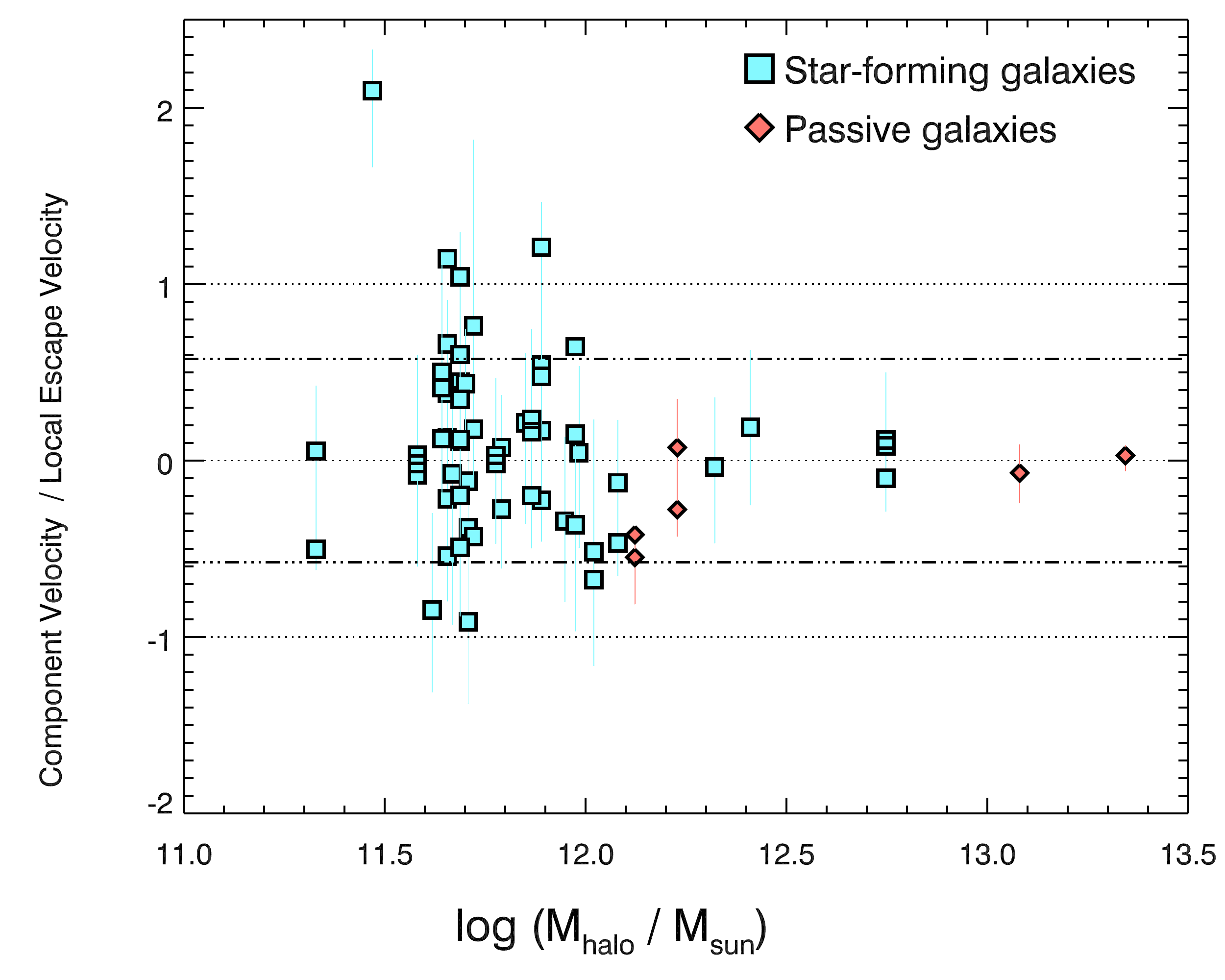}
\label{fig_halos_vratio_white}
\end{center} 
{\bf Figure S1:} O VI component velocities relative to local escape velocity. The dotted line marks a ratio of unity, corresponding to the local escape velocity, and the dot-dashed line marks the relative galaxy/absorber velocity divided by $\sqrt{3}$ times the local escape velocity. 
\end{figure*} 

To compare these velocities with the escape velocities of the halos in which they reside, we convert the measured stellar masses to total dark-matter halo masses using the relation of \cite{Moster:10:903}. We then assume a spherically symmetric NFW profile ($c = 15$) and calculate the escape velocity as a function of halo mass at $R$ = 50, 100, and 150 kpc (Figure 2 of the main text). The absorption does not tend to significantly exceed the estimated escape velocities of the galaxies. This is shown by the cyan range bars on the data points in the lower panel, which mark the velocity range over which each absorber was integrated (e.g. full width at zero optical depth). Some absorbers have their full ranges touch or slightly exceed $v_{\rm esc}$, but this is generally attributable to line wings. 

Since we observe the detected material projected on the sky, we measure its velocity relative to the galaxy projected onto the line of sight instead of its true space motion. Thus the measured relative velocities are strictly lower limits to the relative velocity between galaxy and absorbing cloud. To assess the importance of this effect, we show in Figure S1 the relative galaxy/absorber velocity data shown in Figure 2 of the main text, now divided by its {\it local} escape velocity (for its $R$). Thus any material with a line-of-sight velocity exactly equal to $v_{\rm esc}$ will fall on the dotted lines at $\pm 1$. To account for unconstrained projection effects, we plot also the relative galaxy/absorber velocity divided by $\sqrt{3} v_{\rm esc}$, where the prefactor comes from the mean correction between three-space velocity and the line-of-sight velocity for a random velocity field. Even with this correction, most of the detected absorption falls within the escape velocity. We therefore conclude that we have no strong evidence that most of the detected high-ionization gas is escaping or has escaped its host galaxy halo. 

There remains the chance that these galaxies have associated O VI absorption at higher relative velocities that remains undetected because it falls below our S/N limits. Roughly, we can estimate that this absorption would have a factor of $\sim 10-30$ less column density than the detected absorption, since the detections typically have column densities that are $10-30\times$ the detection limits. \\

\noindent{\Large \bf S5 \,\, O VI and Trends with Star Formation} \\

We find a sharp difference in the detection rate and typical strength of O VI absorption surrounding galaxies that are star-forming and galaxies that are passively evolving. This trend is clearly evident in Figures 2 and 3 of the main text, where we use specific star formation rate as the key variable, because it cleanly divides those galaxies where we have detected star formation from those where we have not, at sSFR $\approx 10^{-11}$ yr$^{-1}$. For the 30 galaxies in which we have detected star formation (those with sSFR $>10^{-11}$ yr$^{-1}$), there are 27 detections of O VI (binomial probability $P = 0.90\pm0.07$ at 95\% confidence), while for the 12 galaxies with lower sSFR there are 4 O VI detections ($P = 0.33 \pm 0.27$ at 95\% confidence). 

We can also test the statistical significance of the apparent correlation of O VI with sSFR. We use a generalized Kendall's tau statistic that can handle doubly censored data \cite{Isobe:86:490}. First, we establish that the correlation of O VI with star formation (sSFR) is strong, when the passive galaxies are included. For all 42 sample galaxies, we obtain $\tau = 0.617$ (significance $P = 1.32\times10^{-5}$) and so can reject at $>99.99$\% confidence the null hypothesis that there is no correlation between $N_{\rm OVI}$ and sSFR. This no-correlation null hypothesis is rejected at $>98$\% in the inner two annular bins of $R$ (Figure 1 in the main text), and at 93\% over $R = 100-160$ kpc. Thus there is strong evidence that the correlation with the presence or absence of star formation holds at all radii in the detected halos. 

The well-known correlation of galaxy star formation with stellar mass is potentially a confusing influence on the O VI, one which hinders our ability to conclusively identify star formation as the key influence on halo O VI. However, even though the size of our sample is relatively small, we have evidence that star formation itself, and not simply stellar mass, has a direct relationship with the observed O~VI. To assess this case we consider the star-forming vs. passive comparison in the range of stellar mass where we have both types of galaxies. We take subsamples with $\log M_* > 10.5$, where there are 11 star-forming galaxies (8 detections) and 12 passive galaxies (4 detections). 

We first consider how the quantity of observed O VI differs between star-forming and passive galaxies in the same mass range. This comparison appears in Figure S2, where we repeat the $N_{\rm OVI}$ versus sSFR plot shown in the upper panel of Figure 3 in the main text but this time including only those galaxies with $\log M_* > 10.5$. Here we see that the strong difference in hit rates between star-forming and passive galaxies is attributable to the very different column density distributions that they follow. The detections of O~VI surrounding passive galaxies are generally weaker than the detections in the star-forming sample, while the non-detections surrounding passive galaxies imply weaker absorption than almost all the detections in either category. If we compare the star-forming and passive galaxies in terms of $N_{\rm OVI}$ using a two-sided Kolmogorov-Smirnov test, we find that we can reject at $>99$\% confidence the null hypothesis that the two column density subsamples are drawn from the same parent distribution (KS statistic $D = 0.644$, probability $0.00866$). 

As a consequence of the lower typical O VI column density surrounding passive galaxies, their hit rate is lower than for the star-forming galaxies. We can use the Wilson score interval to estimate the distribution of the underlying binomial probability of detection, given an observed sample size and hit rate. For the star-forming galaxies, the underlying binomial probability of a detection lies in the range $0.36 - 0.93$ at 99\% confidence with a hit rate of 0.73. For passive galaxies, the 99\% confidence interval is $0.36 - 0.68$ with a hit rate of 0.33. Thus, the hit rate for each category lies just outside the 99\% confidence interval on the binomial probability given by the other, showing that the two rates are different at $\sim 2.6 \sigma$. Because of the lower column densities and hit rates for passive galaxies, even when we control for stellar mass, we conclude that the dichotomy of O~VI is related to star formation, not solely to mass. We still lack the statistics necessary to identify what fraction of the variation in O~VI between the blue cloud and the red sequence is related to star formation and what fraction owes to stellar mass.

\clearpage
\begin{figure*}[!ht]
\begin{center} 
\includegraphics[width=4.5in]{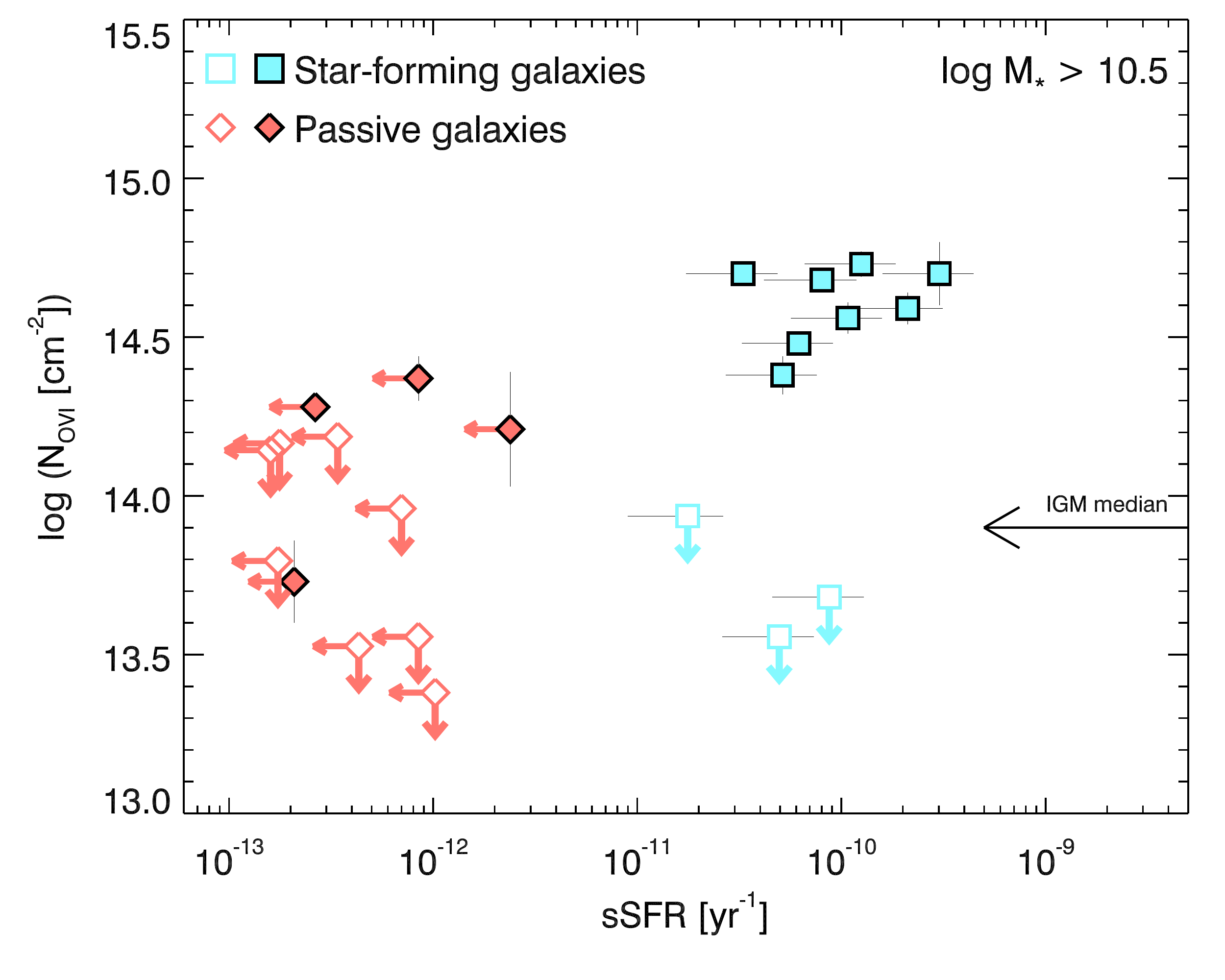}
\label{fig_colden_ssfr_mcut}
\end{center} 
\end{figure*} 
{\bf Figure S2:} O~VI vs. sSFR for only those galaxies with stellar masses $\log M_* > 10.5$.  This figure includes all 12 passive galaxies and 11 of the star-forming galaxies from the main sample. The column density distributions of O~VI are clearly different even over the stellar mass range that star-forming and passive galaxies have in common.  
\vspace{0.3in}


From inspection of Figure~3 in the main text and Figures~S2 and S3 here, it is evident that much of the statistical power of the full sample resides in the red galaxies with no detected star formation and little O VI. We can also test whether the O VI is influenced by the quantity of star formation, or only its presence or absence. To do this we omit the passive galaxies from the sample and use only the 30 galaxies for which star formation was detected. 

Removing the SFR upper limits weakens these correlations significantly. For the three 50 kpc annular bins the no-correlation null hypothesis can only be rejected at $\sim 75 - 80$~\% confidence (Spearman's $\rho  \sim 0.37-0.39$, $P \sim 0.21 - 0.26$). However, for the whole dataset over $R = 0 - 160$ kpc, we still see an indication ($\rho = 0.352$, $P = 94$\% confidence with $N = 30$) of a correlation of O VI with sSFR. Though these correlations are less statistically significant than those for the whole sample including passive galaxies, there is an indication that the quantity of O VI is at least weakly dependent on the quantity of nearby star formation and not simply to its presence or absence. 

\clearpage
\begin{figure*}[!ht]
\begin{center} 
\includegraphics[width=4.5in]{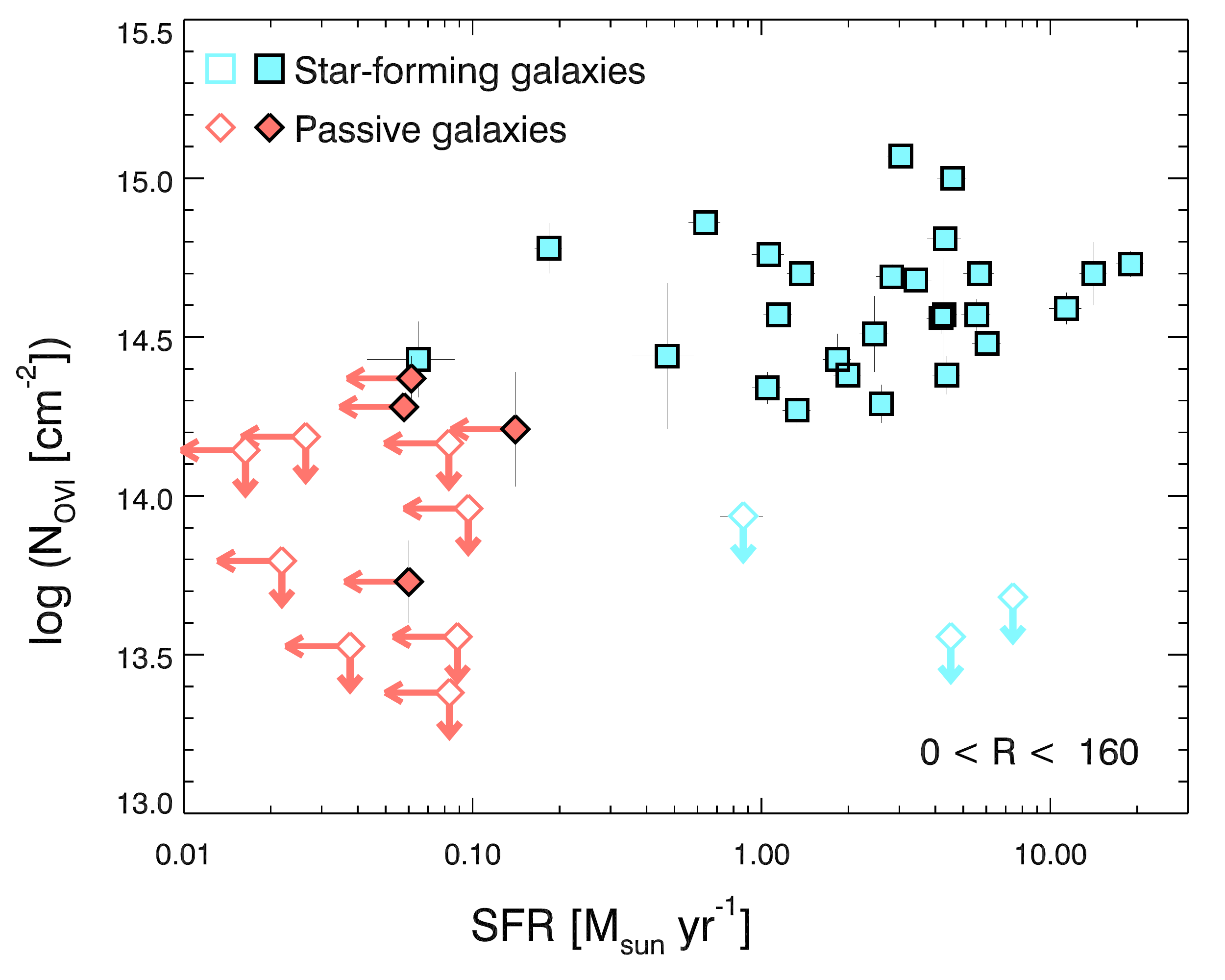}
\label{fig_colden_ssfr_rcuts}
\end{center} 
\end{figure*} 
{\bf Figure S3:} O VI vs. SFR over all $R$ in the sample out to 160 kpc.
\vspace{0.3in} 

Previous characterizations of gas in galaxy halos at low redshift have not found a strong correlation between halo gas and galaxy bimodality. The behavior of O~VI in halos contrasts with that of the cold gas clouds traced by singly-ionized magnesium, Mg II ({\it 30}), which is rarely detected at $R > 100$ kpc, has a steeper radial dependence within that radius, and has a positive correlation with galaxy luminosity that is opposite to our findings for O~VI. Our findings on O~VI are in better accord with the $140$ kpc size of the triply-ionized carbon (C IV) envelopes surrounding galaxies as reported by \cite{Chen:01:158}, but that study did not detect a correlation with star-formation (as proxied by morphological type). Thus the O~VI, C IV, and Mg II may trace distinct phases of the circumgalactic medium with possibly different origins. Studies of star-forming ``Lyman-break'' galaxies at $z \sim 2-3$ have found a connection between star-formation and outflowing material traced by Ly$\alpha$ and low and intermediate ions ({\it 6,7}), and total masses of cold gas comparable to that in the galaxies themselves. \\

\noindent{\Large \bf S6 \,\, Physical Conditions, Distribution, and Mass of the O VI-bearing Halo Gas} \\

The narrow range of physical conditions in diffuse gas that contains significant O VI allows us to obtain robust lower limits to the mass of oxygen in the detected halo gas. O VI is not the dominant ion of oxygen in astrophysical conditions, so the ionization correction applied to derive the total oxygen density (or mass) from the O VI density (or mass) is always substantial (Figure 4 of the main text). To estimate the correction, we have considered the ionization state of O VI over a wide range of temperature as computed by the CLOUDY photoionization code \cite{Ferland:98:761} assuming ionization equilibrium and including both collisional ionization and photo-ionization with the $z=0.2$ radiation background from \cite{Haardt:01}.  Under plausible CGM physical conditions, the ionization timescales are usually $10^7-10^8$ years or less, compared to halo dynamical timescales of $\sim 10^9$ years, so it seems unlikely that a large fraction of the CGM gas could be far from ionization equilibrium.
 
To calculate the mass implied by our detection of strong O VI surrounding star-forming galaxies, we can exploit the fact that column densities are not sensitive to the exact distribution of densities along the line of sight. We can therefore apply basic geometry to deduce the total mass of O VI in a volume projecting a circular region of $R = 150$ kpc on the sky with a typical column density $\log N_{\rm OVI} = 14.5$. The total mass of O VI in solar masses is: 
\begin{equation}
M_{\rm OVI} = \pi R^2 N_{\rm OVI} (16 m_H).
\end{equation} 

Scaling to typical values for our sample and including the O~VI ionization correction, we obtain the total oxygen mass: 
\begin{equation} 
M_{\rm O} = 1.2\times10^7 M_{\odot} \left( \frac{R}{150 \,{\rm kpc}} \right)^2 \left( \frac{N_{\rm OVI}}{10^{14.5 }\, {\rm cm}^{-2}} \right) \left( \frac{0.2}{f_{\rm OVI}} \right),
\end{equation} 
which has the same functional form and scalings as the relation in the main text. As in the main text we also correct the coefficient from 1.4 down to $1.2$ according to the measured covering fraction around star-forming galaxies in three radial bins: 11/11 detections with $R < 50$ kpc, 9 of 10 with $50 < R < 100$, and 7/9 with $100 < R < 150$. Taking the mean column density separately in each of these three bins gives $\log \langle N_{\rm OVI} \rangle =$ 14.7, 14.6, and 14.5 respectively. Adopting the $\langle N_{\rm OVI} \rangle$ and $f_{hit}$ for the individual bins gives a mass coefficient $1.4 \times 10^7 M_{\odot}$, but to impose a firm lower limit for the main text we have adopted the lower envelope of the star-forming detections at $\log \langle N_{\rm OVI} \rangle = 14.5$ to avoid being skewed by the few high values at small $R$. 

The interstellar gas masses cited in the text were derived from the individual stellar masses by applying the mean gas fraction versus stellar mass relation as tabulated by ({\it 27}). The typical interstellar oxygen masses in the galaxies were then obtained by applying the metallicities, $Z = 0.5 - 1.2 Z_{\odot}$, obtained from the measurements of nebular emission lines. These values define the green band in Figure 4 of the main text.

\clearpage
\begin{figure*}[!t]
\begin{center} 
\includegraphics[width=4.5in]{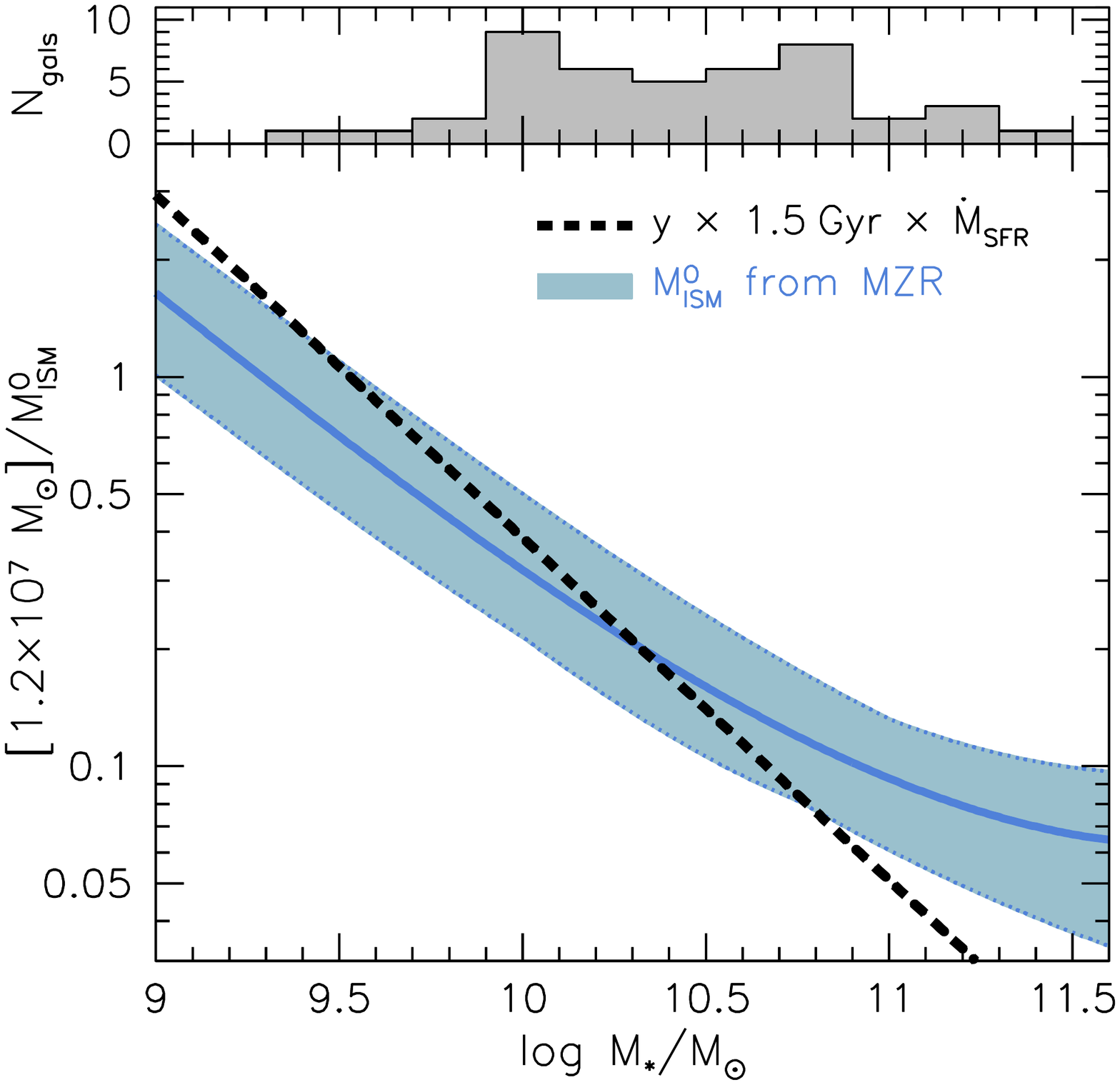}
\end{center} 
\label{oxy_frac_fig}
\end{figure*} \vspace{-1.5in}
{\bf Figure S4:} The ratio between our inferred minimum CGM oxygen mass ($M_{\rm O} = 1.2 \times 10^7 M_{\odot}$, see Equation 2) and the typical ISM oxygen mass $M_{\rm ISM}^{\rm O}$, as a function of galaxy stellar mass, and with the conservative O VI ionization correction $f_{\rm OVI} = 0.2$. The blue curve and band correspond to the mean observed MZR and its systematic uncertainties. The CGM oxygen mass $M_{\rm O}$ range from 70\% of $M_{\rm ISM}^{\rm O}$ at the lower end of the mass range in our sample, $10^{9.5} M_{\odot}$, down to 10\% at upper mass limit of $10^{11} M_{\odot}$. Because the CGM oxygen mass increases in inverse proportion to $f_{\rm OVI}$, the blue band expresses a lower limit to the CGM-to-ISM ratio. The heavy dashed line shows the ratio between the inferred minimum CGM mass and the oxygen mass produced by these galaxies over the last 1.5 Gyr. 
\vspace{0.3in}

We have also calculated $M_{ISM}^{\rm O}$ for typical star-forming galaxies, using a mean gas fraction and mean mass-metallicity relation for SDSS galaxies following the analysis of ({\it 27}). These results are shown in Figures 4 and S4. Here we have derived the gas fractions of galaxies using the total budget of cold gas (atomic plus molecular). The mean ISM oxygen mass as a function of galaxy stellar mass is then derived and compared with the minimum halo oxygen mass, $1.2 \times 10^7 M_{\odot} \times (0.2  / f_{\rm OVI})$. The widths of these hashed regions express the systematic uncertainty in the mass-metallicity relation.  For the mean mass-metallicity relation adopting the total cold gas budget, we find that $M_{\rm O} / M_{\rm ISM}^{\rm O}$ = 0.7 at $\log M_* = 9.5$ and 0.1 at $\log M_* = 11$ (Figure~S4). This ratio increases as $f_{\rm OVI}$ declines from its maximal value of 0.2. Notably, the ratio between CGM and ISM oxygen masses increase as galaxy stellar mass declines, if $f_{\rm OVI}$ is held fixed. 

We can also consider the ratio between the measured CGM oxygen mass $M_{\rm O}$ and the oxygen produced by star formation over some timescale. In this comparison we adopt the mean SDSS SFRs as a function of stellar mass, the oxygen yield of 0.014 $M_{\odot}$ per $M_{\odot}$ of star formation, and a timescale of 1.5 Gyr, the travel time to 150 kpc of a cloud moving at 100 km s$^{-1}$. The minimum ratio between the detected CGM oxygen mass and this ``produced'' mass is given by the heavy dashed line in Figure S4. As stated in the main text, the detected CGM mass is a notable fraction of the total amount of oxygen produced by these galaxies over a significant timescale. 

Because we measure column densities, many possible arrangements of gas along the line of sight are possible. However, we can estimate the total line-of-sight extent of the gas using simple scalings. For a solar ($Z=Z_\odot$) oxygen abundance ({\it 26}), the O VI column density is
\begin{equation}
N_{\rm O VI} = L \left({n_{\rm O}\over n_{\rm H}}\right)\left({\rho \over m_{\rm H}} \right) \left({ Z  \over Z_{\odot} }  \right)
 = 10^{14} L_{\rm kpc} \left({\rho/\bar{\rho}\over 1000}\right)
           \left({ f_{\rm O VI}  \over 0.2 }\right)  \left({ Z  \over Z_{\odot} }  \right) \,{\rm cm}^{-2}~,
  \label{a}
\end{equation}
where $L_{\rm kpc}$ is the path length in kpc and $\bar{\rho}$ is the mean cosmic density at $z = 0.2$ (particle density of $n_H = 3.26 \times 10^{-7}$ cm$^{-3}$). Thus halo clouds with density $n_H = 10^{-4}\,{\rm cm}^{-3}$ and $f_{\rm OVI}=0.2$ can give $\log N_{\rm OVI} = 14.5$ if they extend 10 kpc along the line of sight, but the required path length rises to 200 kpc if the ionization fraction of O VI is only 1\% rather than its peak value of 20\% (Figure S5). Except at very low densities, where the cloud sizes become too large to fit within galaxy halos and still give the column densities we observe for star-forming galaxies, the ionization fraction of O VI is never more than $f_{\rm OVI} = 0.2$. 

The cited oxygen masses for the Mg II halos of galaxies were obtained by taking the total baryonic mass estimate for that medium from ({\it 30}), correcting down by 0.1 solar metallicity as argued in that paper, and then applying the solar relative abundances of magnesium and oxygen. The cited mass estimate for Galactic HVCs is taken from the estimate by ({\it 28}) for the more conservative scenario that locates all the HVCs within 60 kpc of the Milky Way disk. 

To calculate the fraction of cosmic oxygen represented by the ionized halos of star forming galaxies, we apply the generic $R = 150$ kpc halo with $N_{\rm O VI} = 10^{14.5}$ cm$^{-2}$ to all star-forming galaxies with $M_* > 10^{9.5} \, M_{\odot}$, the range covered by our survey. Integrating over the representative population (specified by the K-selected late-type luminosity function of \cite{Bell:03:289}, with $\alpha=-0.94$), we obtain $\Omega _{\rm OVI}^{\rm halos} = 8.26/h \times 10^{-8} \times (1/ f_{\rm OVI})$. The cosmic density of oxygen is $\Omega _{\rm O} = y\Omega_* = 0.014 \times 0.00197/h = 2.76/h \times 10^{-5}$. The fraction of cosmic metals in star-forming halos like those in our sample is then $\Omega_{\rm OVI}^{\rm halos} / \Omega_{\rm O} = 0.015 \times (0.2/ f_{\rm OVI}) $, as given in the text.  If we correct this cosmic mass budget to the total gas mass implied in these halos, scaling to the Milky Way HVC abundances, we obtain $\Omega _{b} ^{\rm halos} = 177 (Z_{\odot} / Z) \Omega_{\rm OVI}^{\rm halos} = 7.31 / h \times 10^{-4} \times (0.1Z_\odot / Z) \times (0.2 / f_{\rm OVI})$. Dividing by the cosmic baryon fraction \cite{Dunkley:09:306}, we obtain $\Omega _{b} ^{\rm halos} = 0.023 (Z_\odot / Z) \times (0.2 / f_{\rm OVI})$. Thus the ionized gas in the extended halos of our star forming sample represent at least 0.2\% of cosmic baryons if they have $f_{\rm OVI} = 0.2$ and solar metallicity, 2\% if conditions are not optimized for O~VI ($f_{\rm OVI} \sim 0.02$), or if the metallicity is $0.1 Z_{\odot}$, or up to 20\% if they have both low metallicity and low $f_{\rm O VI}$. The contribution of these halos to the baryon budget is similar to or exceeds that from atomic and molecular phases of the ISM \cite{Fukugita:98:503}, which also make $1-2$\% contributions, but only if conditions are not optimal for O VI and/or the gas metallicity is low. 

Figure S5 suggests that it is just possible to explain our observed column densities with solar metallicity, photo-ionized gas at $T \leq 10^5\,$K, provided that this gas fills enough of the halo to accumulate path lengths $\geq 100\,$kpc.  However, for $Z=0.1 Z_\odot$ the required path lengths become much larger than the virial radii of the galaxy halos, expected to be $200-350\,$kpc for galaxies in our observed stellar mass range.  Furthermore, if halo gas traces the dark matter, then typical overdensities at radii $\sim 100\,$kpc are several hundred or more, in which case photo-ionization is ineffective.  These arguments favor collisionally ionized gas, which requires finely tuned temperatures to achieve high $f_{\rm OVI}$ (see Figure~4 and discussion in the main text). There are two significant caveats to the argument for collisional ionization.  First, the gas responsible for the OVI absorption could have lower overdensity than the halo dark matter, with $(\rho/\bar{\rho})_{\rm gas} \sim 50-100$ at $R \sim 100\,$kpc. Large path lengths like those shown in Figure S5 would still be required, so this solution is only viable if the metallicity is close to solar. Second, the UV background at the 10 Rydberg ionization threshold of O VI could be a factor of several higher than the value we have adopted \cite{Haardt:01}, in which case photo-ionization becomes important at an overdensity higher by the same factor. To give a concrete example, if the UV background intensity is a factor of four higher, then $f_{\rm OVI} \approx 0.05$ at $T=10^{4.5}\,$K and $\rho/\bar{\rho}=400$ (see the $\rho/\bar{\rho}=100$ curve in Figure~4), and the required path length for $N_{\rm OVI}=10^{14.5}\,{\rm cm}^{-2}$ drops from $100(Z_\odot/Z)\,$kpc (Figure S5, red curve) to $25(Z_\odot/Z)\,$kpc.  With our current data, we cannot determine the metallicity of the absorbing gas or decide between collisionally ionized and photo-ionized scenarios (or a mix of the two).  In any case, however, achieving our high covering factors and observed column densities requires that favorable combinations of physical conditions arise in the typical halos of star-forming galaxies.

\clearpage 
\begin{figure*}[!ht]
\begin{center} 
\includegraphics[width=4.5in]{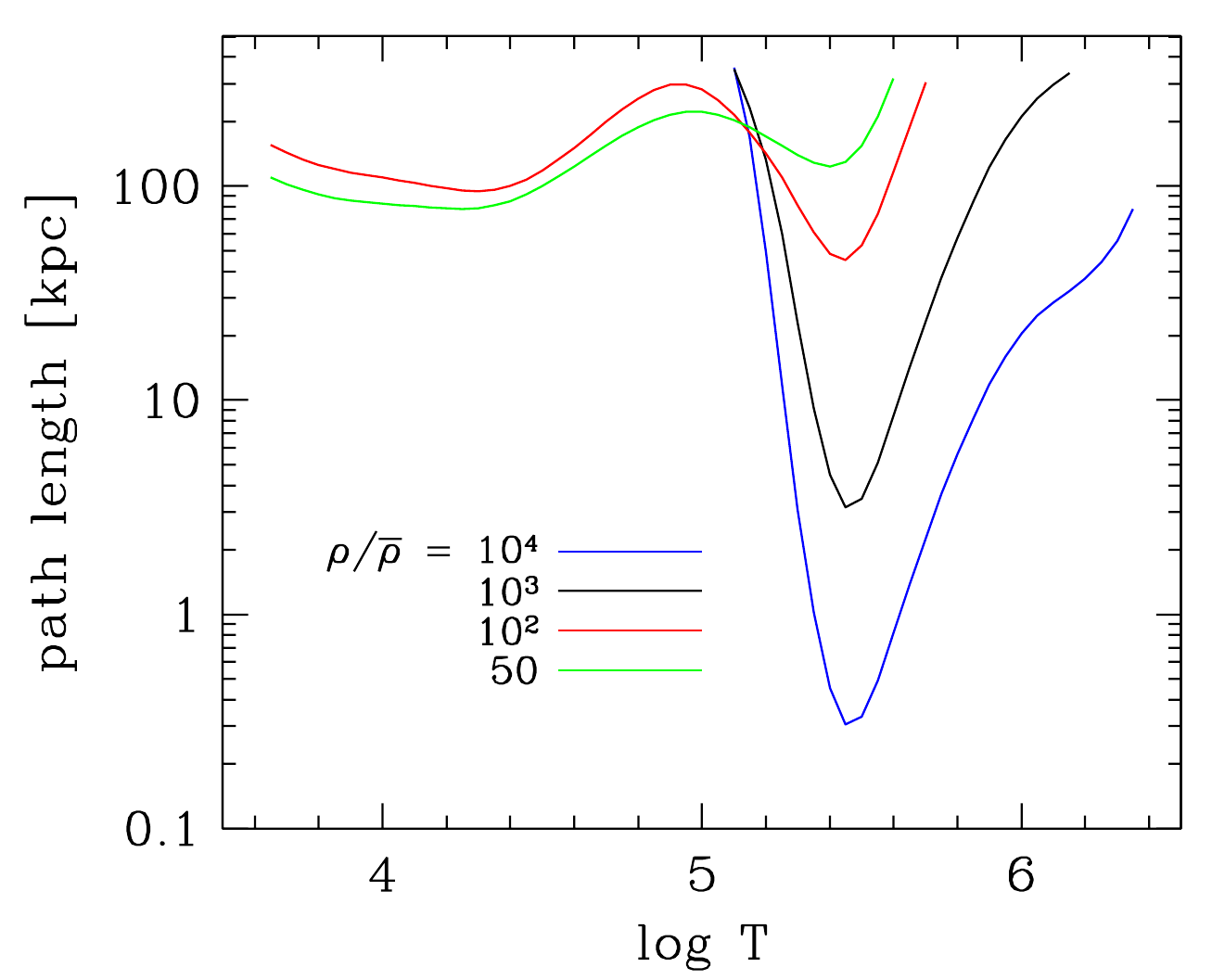}
\end{center} 
\label{ovifracfig}
\end{figure*} 
\vspace{-0.2in} 
{\bf Figure S5:} Path length required to obtain $N_{\rm OVI} = 10^{14.5}\,{\rm cm}^{-2}$ for solar metallicity gas with the indicated overdensity and temperature, using the $f_{\rm OVI}$ values shown in Figure~4 of the main text.  These pathlengths scale in inverse proportion to metallicity according to Equation (3), and so are $10\times$ longer for $Z = 0.1 Z_{\odot}$.  The overdensity and ionizing background are computed at $z=0.2$, using the UV background model of \cite{Haardt:01}. Gas with $\rho/\bar{\rho} \geq 1000$ is close to collisional ionization equilibrium, so $f_{\rm OVI}$ depends only on temperature and required path lengths are inversely proportional to density.  At overdensity $\rho/\bar{\rho} \leq 100$, photo-ionization becomes important for $T < 10^{5.5}\,$K, so $f_{\rm OVI}$ depends on the gas density and the scaling of required path length with density is no longer simple.
\vspace{0.3in}

\bibliographystyle{Science}

\begin{thebibliography}{40}
\bibitem{metals} In astronomical usage, metals are those elements heavier than hydrogen and helium that are formed only by stellar nucleosynthesis.
\bibitem{OVI_igm_note} The typical $\log N_{\rm OVI} = 14.5 - 15.0$ for star-forming galaxies resembles the high end of the column-density distribution seen in blind surveys of intergalactic clouds \cite{Danforth:08:194, Tripp:08:39, Thom:08:22} and is higher than the mean value (14.0) measured by the Far Ultraviolet Spectroscopic Explorer through the halo of the Milky Way \cite{Sembach:03:165}, which would belong in our star-forming sample.
\bibitem{OVI_ell_note} O~VI {\it emission}  is seen in elliptical galaxies \cite{Bregman:05:1031}, but this gas is most likely associated with the ISM and not the CGM.
\end{thebibliography}

\end{document}